\documentclass[aps,twocolumn,showkeys]{revtex4}
\usepackage{epsfig}
\usepackage{amsmath}
\usepackage{cases}
\usepackage[usenames]{color}

\begin{document}

\date{\today}

\author{F. Perakis, M. Mattheakis and G. P. Tsironis}

\affiliation{Crete Center for Quantum Complexity and Nanotechnology, Department of Physics, University of Crete, P.O. Box 2208, 71003 Heraklion, Greece}

\affiliation{ Institute of Electronic Structure and Laser, Foundation for Research and Technology-Hellas, P.O. Box 1527, 71110 Heraklion, Greece}

\title {Small-world networks of optical fiber lattices}

\begin{abstract}

We use a simple dynamical model and explore coherent dynamics of wavepackets in complex networks of optical fibers. We start from a symmetric lattice and through the application of a  Monte-Carlo criterion  we introduce structural disorder  and deform the lattice into a small-world network regime. We investigate in the latter both structural (correlation length) as well as dynamical (diffusion exponent) properties and find that both exhibit a rapid crossover from the ordered to the fully random regime. For a critical value of the structural disorder parameter $\rho \approx 0.25$ transport changes from ballistic to sub-diffusive due to the creation strongly connected local clusters and channels of preferential transport in the small world regime.

\end{abstract}

\maketitle

\section{Introduction}

Small world networks have attracted great interest in the last decades due to the very broad applicability to real-life problems ranging from social networks, to physics, chemistry and biology~\cite{strogatz_exploring_2001,vendruscolo_small-world_2002,newman_structure_2003,rao_protein_2004,pomerance_effect_2009,kempes_predicting_2011}. Some of the unique properties that arise in the small-world regime are robustness (high clustering coefficient), combined with efficient transport properties (low average path length) ~\cite{watts_collective_1998,liu_controllability_2011,halu_2013}. Quantum dynamics simulations on scale free networks indicate a phase transition in the transport properties as one approaches the thermodynamic limit~\cite{halu_2013,halu_2012, bianconi}. Even though small-world networks have been extensively studied focusing on diffusive properties and using a classical statistical description, wave-like propagation properties have not attracted considerable attention yet.
In the present contribution, the small-world concept is applied to an optical or  equivalently to a quantum mechanical system. In particular we model an optical fiber lattice that exhibits small-world network topology. The transport properties of this network are investigated and compared with those of an ordered as well as a uniformly random network.  The dynamics of the system are governed by the amount of structural disorder present in the network.

Experimental investigations of disordered optical lattices indicate that a small amount of disorder is sufficient to lead to Anderson localization~\cite{schwartz_transport_2007,lahini_observation_2009,segev_anderson_2013,naether_enhanced_2013,vardeny_optics_2013}.  In this case the excited wavepacket does not disperse but instead remains partially localized due to the dielectric index disorder in the medium. A different source for localization is nonlinearity:  when the interaction of the wavepacket with the medium is highly non-linear it can lead to self-trapping~\cite{hennig_wave_1999,li_self-trapping_2013}. For example, when a laser pulse interacts with a highly responsive medium, it can lead to self focusing and soliton formation~\cite{picozzi_solitons:_2008,gao_iterative_2010,mattheakis_luneburg_2012}. Self-trapping is known to occur abruptly at a threshold which depends on the system properties such as size and geometry~ \cite{molina_dynamics_1993,tsironis_applications_1993,molina_disorder_1995,perakis_discrete_2011,tsironis_exact_2011}. 

In the present article we investigate primarily the impact of structural disorder on the wavepacket dynamics on a 2D lattice.  The specific form of disorder used leads to the formation of a small world network lattice that is parametrized through a randomness parameter.  Additionally, we observe the effects that nonlinearity may induce, viz. that nonlinearity rapidly leads to self-trapping, while structural disorder acts like a barrier creating channels in which transport is favoured, leading to anomalous diffusion. Previous investigations on optical random media indicate that nonlinearity can emerge from structural disorder in the form of branched wave patterns~\cite{ni_origin_2011}. In the current investigation we essentially disentangle these two parameters and explore the possibility of tuning independently localization due to structural disorder and due to self-trapping.

\section{The small-world lattice}
\begin{figure*}[tbh]
	\begin{center}
	\includegraphics[trim = 20mm 110mm 20mm 5mm, clip,scale=0.9]{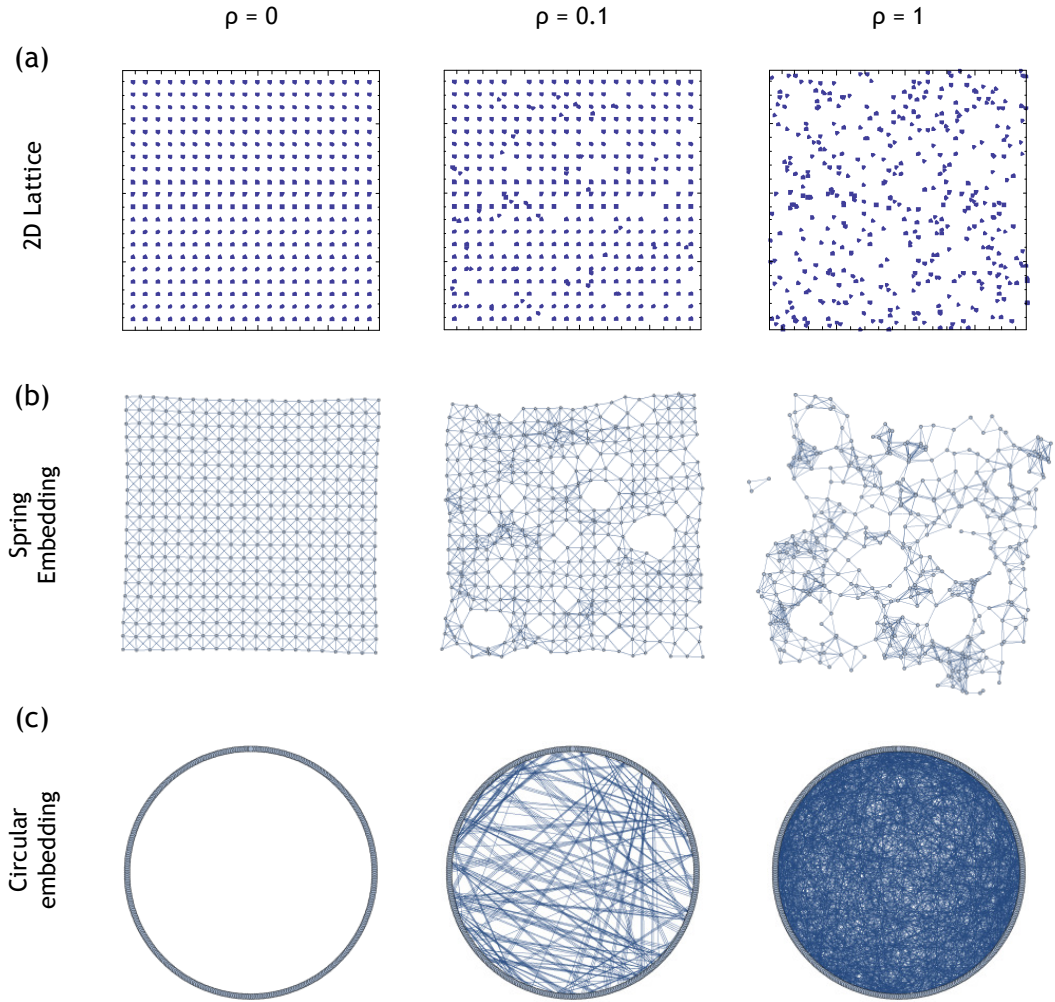}
	\caption{(a)The 2D lattice of the optical fiber position for different values of the structural disorder parameter $\rho$. The corresponding network configurations in the (b)spring  and (c)circular embedding.  }
	\label{lattice_network}
    \end{center}
\end{figure*}

The physical system we are adressing consists of a lattice or network of optical fibers that permits light propagation along
the fibers plus some interfiber interaction due to evanescent coupling.
In Fig.\ref{lattice_network} we show an ordered square lattice, where the points correspond to the optical fibers position on the x-y plane. The fibers extend along the z direction, which in a paraxial approximation corresponds to time~\cite{marte_paraxial_1997}. Initially, one of the fibers is excited and by measuring the intensity profile of the lattice for different cuts along the z-axis, one acquires the dynamics of the excited wavepacket on the lattice.

Starting from an ordered lattice, structural disorder is inserted to the system by applying a Monte-Carlo criterion. A disorder parameter $\rho$ is introduced, which is the probability of moving a single point of the lattice to a new position. Sampling the whole lattice, the points are repositioned   by decision (whether a random check between 0 and 1 is larger that $\rho$) to a new random position in the plane. The case $\rho=0$ is  the fully ordered case, while for $\rho=1$ all the points have been repositioned at arbitrary positions yielding a uniform random network. In the intermediate regime $0<\rho<1$ the lattice is partially disordered; we will show that  a  small amount of structural disorder is sufficient for the emergence of the small-world properties.

We can construct a network out of a 2D lattice by assigning the fibers as the network nodes while the interaction between the optical fibers are the edges. This treatment yields a weighted network~\cite{barrat_architecture_2004}, where the weight of the edges depends on the distance between the points on the 2D plane in the following way:

\begin{eqnarray}
w_{ij} = A\cdot e^{-(r_{ij}-r_0)}
\end{eqnarray}
where $w_{ij}$ is the weigth that links any two fibers labeled $i$ and $j$ respectively, $r_{ij}$ is the distance between 
these fibers, $r_0$ a characteristic lenght scale and $A$ is weight amplitude.
This form of edges models reasonably well the fiber crosstalk, i.e. the tunneling probability between fibers. In Fig.\ref{lattice_network}b we show the networks that arise from the 2D lattices for the cases where $\rho=0$, $\rho=0.1$ and $\rho=1$. To facilitate comparison between the three networks the edges are shown only above a certain weight (0.01$\%$). The algorithm used to optimize the network topology is the spring embedding, which assumes a ball-string relationship between the nodes of the system.

\begin{figure}[bth]
	\begin{center}
	\includegraphics[trim = 20mm 205mm 100mm 5mm, clip,scale=0.9]{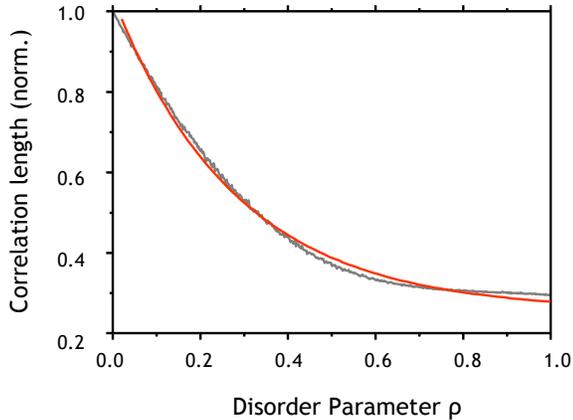}
	\caption{The correlation length of the network as a function of the disorder parameter $\rho$. A exponential fit (red line) yields a characteristic $\rho_l=0.28$.}
	\label{correlation_length}
    \end{center}
\end{figure}

In the ordered case ($\rho=0$) the network is a 2D manifold corresponding precisely to the 2D lattice. For a small amount of structural disorder ($\rho=0.1$) a few nodes have been repositioned, creating holes on the one hand and patches of increased density on the other. This treatment leads to the formation of local clusters, making some of the nodes of the network highly connected and others more disconnected. This is the key feature of the network, which as we will see has an significant impact on the dynamics. In the fully random case ($\rho=1$) the network has been essentially fractured to smaller fractions, which are very weakly connected among them. The exact value that the network passes to the fractured regime relates to the site and bond percolation threshold of the network~\cite{callaway_network_2000,zhao_inducing_2013}.

An more common network graphical representation is the circular embedding, shown in Fig.\ref{lattice_network}c. The nodes are placed on a circle and the corresponding edges are drawn between them. Then instead of repositioning the nodes like in the real 2D lattice and in the spring embedding, they are kept in fixed positions and instead the edges are redirected. From this point of view $\rho$ can also be understood as the network rewiring probability. It has been shown that by following this procedure small-world properties arise at very small $\rho$'s ~\cite{watts_collective_1998}. Using this approach we can analyse the disordered 2D lattice as a weighted network. This method allows us to make a direct connection between the lattice transport properties and the network topology and simplify the simulations by mapping the 2D lattice plane to a reduced adjacency matrix.

\begin{figure}[thb]
	\begin{center}
	\includegraphics[trim = 10mm 205mm 110mm 5mm, clip,scale=0.85]{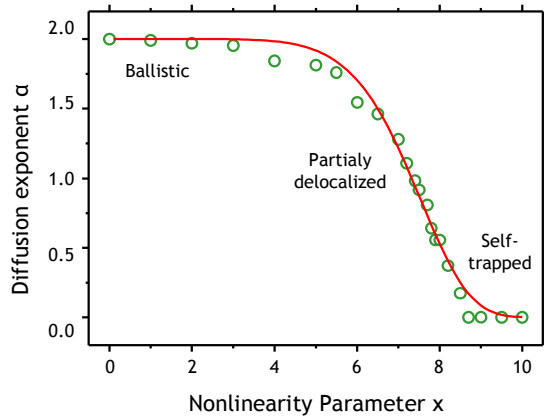}
	\caption{The diffusion coefficient $\alpha$ as a function of the nonlinearity parameter $\chi$. For small $\chi$'s the system exhibits ballistic behaviour, whereas for larger ones ($\chi_c= 7.5$) the wavepacket remains partially localized and finally becomes self-trapped. }
	\label{alpha_chi}
    \end{center}
\end{figure}

\begin{figure*}[thb]
	\begin{center}
	\includegraphics[trim = 20mm 40mm 20mm 5mm, clip,scale=0.9]{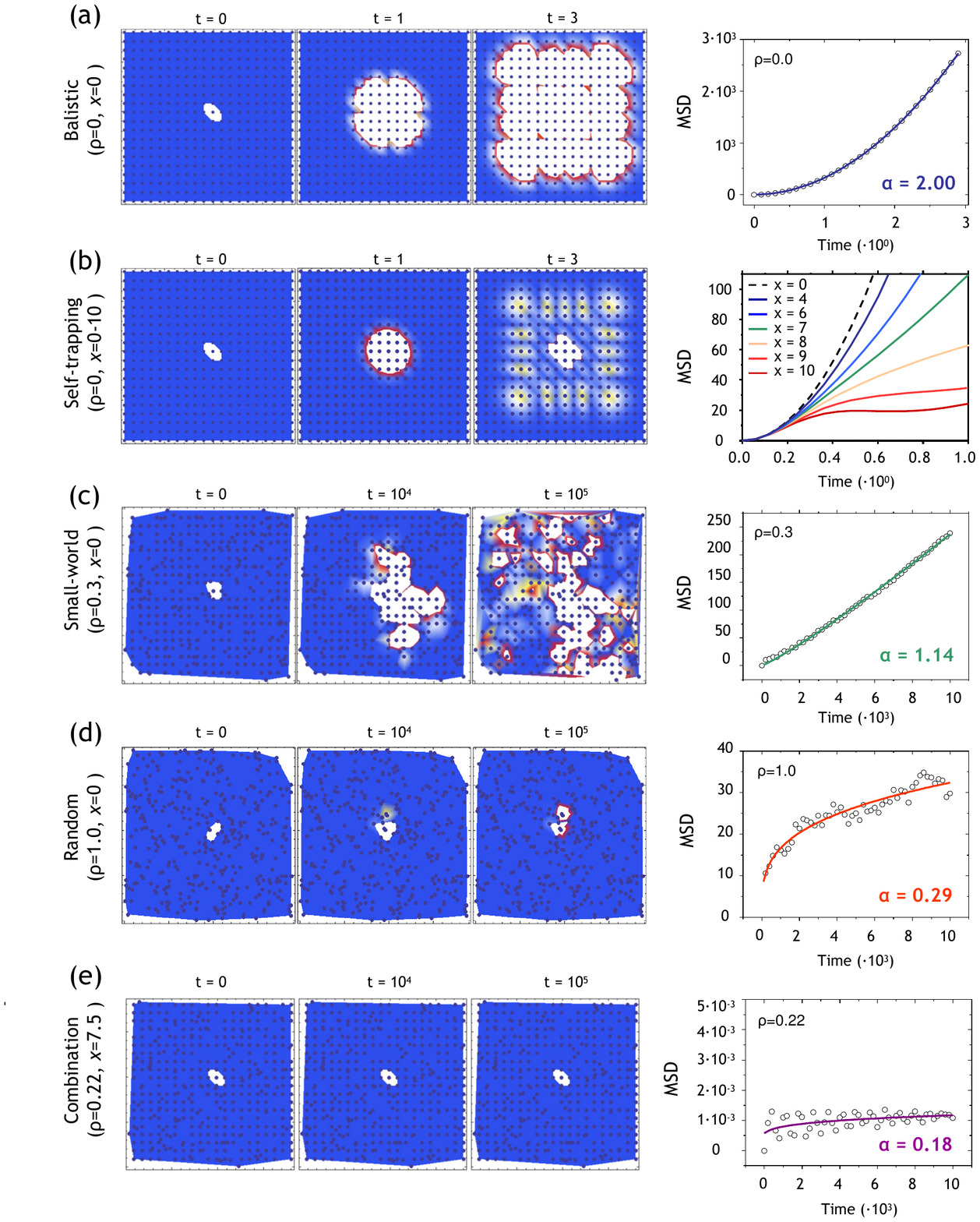}
	\caption{The dynamics of the wavepacket and the corresponding mean--square--displacement (MSD) as a function of time. (a) The linear case in the ordered lattice. (b) Nonlinearity induces self--trapping in the ordered lattice. In presence of structural disorder, the system passes from normal diffusion (c) to sub--diffusion (d). Finally the combination of nonlinearity and disorder leads to strong localization (e).}
	\label{dynamics}
    \end{center}
\end{figure*}

To quantify the structural local change of the lattice as a function of increasing structural disorder, we calculate the correlation length for different $\rho$'s. This can be achieved by calculating the standard deviation of the correlation function of the disordered network, which is normalized with respect to the ordered lattice. In order to have a statistically good ensemble, we make 50 realizations for 500 different steps of the disorder parameter $\rho$.
The resulting correlation length decays rapidly as a function of the disorder parameter, as can be seen from Fig.\ref{correlation_length}.
To extract consistently a single value for the correlation length decay, we fit with a single exponential function (red curve) yielding $\rho_l=0.28$. As we discuss also below this value signals the regime between ordered and random network; this feature emerges in both the structural and dynamical aspects of the system. The corresponding offset of $0.26$,  observed for larger values of the disorder parameter is due to the finite system size (400 points).

\section{Dynamics}

To simulate the dynamics of the excited wavepacket on the 2D lattice use the discrete nonlinear Schr\"{o}dinger equation (DNLS)~\cite{eilbeck_discrete_1985,molina_dynamics_1993,molina_disorder_1995,perakis_discrete_2011}:

\begin{eqnarray}
i\frac{d\psi_n}{dt}=\sum_{m} V_{n,m} \psi_{m}-\chi |\psi_n|^{2} \psi_n
\label{dnls1}
\end{eqnarray}

where V is the network adjacency matrix, which contains the weights between the vertices. The second term introduces anharmonicity with the nonlinearity parameter $\chi$.  The DNLS equation is an ideal choise to depict the evolution, because it can describe realistically the tunneling of a wavepacket through optic fibers and allows one to explore the influence of nonlinearity on the dynamics. The resulting dynamics are displayed on Fig.\ref{dynamics}, where the color values indicate the amplitude of wavepacket between 0 (blue) and 1 (white). Initially the wavepacket is placed in the central lattice site; in an experiment, this initial condition corresponds to the excitation of a single fiber. In the vertical sections we show  selections of snapshots from the evolution of the wavepacket, while on the right hand side we portray the wavepacket mean--square--displacement (MDS) as a function of time. In the ordered lattice with no nonlinearity (Fig.\ref{dynamics}a), where $\rho=0$ and $\chi=0$ the wavepacket spreads ballistically. This is quantified through the function 
 $f(t)=Dt^\alpha$ where D is the diffusion coefficient and $\alpha$ the diffusion exponent. In the ballistic case, the fit  yields $\alpha=2$ (blue line).  We note that these results correspond to finite system behaviour that stop every time light reached the edge fibers due to diffraction.

We also examine the influence of nonlinearity in the ordered lattice (Fig.\ref{dynamics}b). As expected from previous investigations \cite{molina_dynamics_1993,molina_disorder_1995,perakis_discrete_2011,tsironis_exact_2011}, self-trapping of the wavepacket occurs abruptly above a certain value of the nonlinearity parameter $\chi$. This is characteristically depicted on the dynamics, as can be seen from the snapshots for $\chi=10$. Even though the wavepacket initially spreads (t=1), due to the intense self-interaction introduced by the nonlinear term it is then trapped to the initial site (t=3). Additionally, a small fraction of the wavepacket spreads through the rest of the lattice. The mean--square--displacement of the wavepacket is shown for different values of the nonlinear parameter ($\chi=0,..,10$) on the right--hand panel. 

\begin{figure}[thb]
	\begin{center}
	\includegraphics[trim = 10mm 205mm 110mm 5mm, clip,scale=0.9]{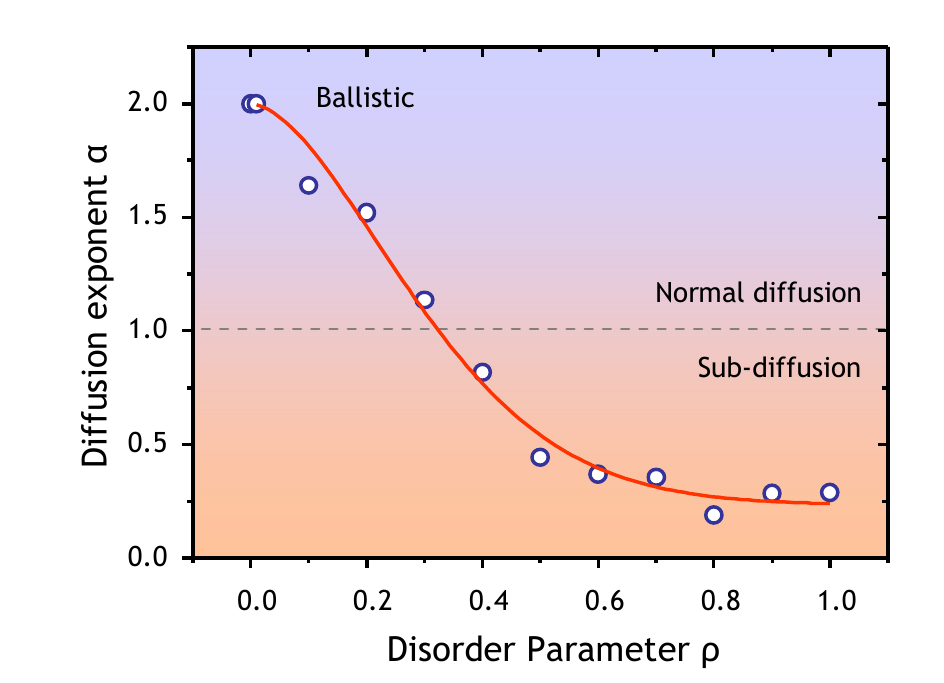}
	\caption{The diffusion coefficient $\alpha$ as a function of the disorder parameter $\rho$. For small $\rho$'s the system exhibits normal diffusion, whereas for larger ones ($\rho_c= 0.22$) it switches to sub-diffusion.}
	\label{alpha_rho}
    \end{center}
\end{figure}

One can see that the wavepacket exhibits ballistic behaviour for smaller values of the nonlinearity parameter, whereas a rapid change of behaviour occurs for greater $\chi$'s. This feature becomes clear when plotting the fitted diffusion coefficients $\alpha$ as a function of the nonlinearity parameter (Fig.\ref{alpha_chi}). When increasing nonlinearity, the system changes  from ballistic behaviour  to partial delocalization and finally to self trapping. A sigmoidal function fit (stretched exponential - red line) yields the characteristic value of $\chi_c= 7.5$, which signifies the passing to the self-trapped regime. This feature has been studied extensively in previous investigations, where it was shown that the value $\chi$ critically depends on the system size and geometry~\cite{molina_dynamics_1993,tsironis_applications_1993,molina_disorder_1995,perakis_discrete_2011,tsironis_exact_2011}. Here, we see that this property also holds for the 2D lattice and additionally discover that a small fraction of the wavepacket is not trapped. In the simulations we use open boundary conditions; we verified that implementing periodic boundary conditions does not significantly alter our results and simply shifts the critical nonlinearity value $\chi_c$ as shown in 1D networks \citep{molina_dynamics_1993}, since we essentially limit our studies on the early wavepacket dynamics, where the probability amplitude is nearly zero near the boundaries. Furthermore, similar effects are observed for the system size: self-trapping is known to depend critically on system size \citep{molina_dynamics_1993}, which again can shift in our system the critical nonlinearity $\chi_c$ in an analogous way to the 1D system.

An alternative way to localize the excitation is by incorporating  disorder, which has been previously observed theoretically and experimentally in optical lattices\cite{schwartz_transport_2007,lahini_observation_2009,segev_anderson_2013,naether_enhanced_2013,vardeny_optics_2013}. Using the procedure described in Fig.\ref{lattice_network}, structural disorder is introduced to the system and the dynamics is  examined in the regime in between an ordered lattice and a completely random one. In Fig.\ref{dynamics}c we show snapshots from the dynamics in a lattice with disorder parameter $\rho=0.3$. Naturally, structural disorder acts like a barrier,  slowing down dramatically all dynamics while creating new channels of transport as can be seen clearly in the latter snapshots ($t=10^5$). As a result the  diffusion coefficient decreases, in this case to $\alpha=1.14$ (the traces shown in the disordered cases are averaged over 50 different lattices). For a completely random lattice (Fig.\ref{dynamics}d - $\rho=1$) we observe that the wavepacket remains localized within the observed timescales and exhibits sub-diffusion with corresponding diffusion coefficient $\alpha=0.29$. This behaviour is quantified in Fig.\ref{alpha_rho}, where is shown the diffusion coefficient $\alpha$ as a function of the disorder parameter $\rho$. We consider this as the main result of the paper: the system passes from normal diffusion to sub-diffusion with increasing structural disorder, even though nonlinearity is absent. A characteristic value of $\rho_c=0.22$ is extracted from the fit (stretched exponential - red line), which signifies the passage to sub-diffusive behaviour. The residual offset ($\alpha_0=0.23$) is due to trapping of the wavepacket in the initial position, which exhibits Anderson-like localization. The transition to sub-diffusion signifies the passage from an ordered network to a small-world regime. Finally including a combination of both nonlinearity and structural disorder leads to very strong localization ($\alpha = 0.18$ ) as shown in Fig.4e for the corresponding critical values $\rho_c$ and $\gamma_c$.

\section{Conclusions}
In the present paper is discussed a simple model describing the dynamics of a wavepacket in an 2D lattice, in the presence of structural disorder and nonlinearity. In an experiment, the 2D lattice can refer to the tips of an optic fiber network; the dynamics examined refer to the excitation of one fiber and recording the intesity of the surrounding ones. 

In an ordered lattice with no nonlinearity, we observe, as expected, that the system exhibits ballistic behaviour. 
Non-linearity is introduced through the DNLS equation, which lead to self-trapping for values larger than $\chi_c= 7.5$. On the other hand, structural disorder is introduced to the system by mapping the 2D lattice to a network, and repositioning the nodes with a Monte-Carlo criterion and probability $\rho$. As structural disorder increases, we observe a transition with regard both the structural and dynamical properties of the system at about $\rho=0.25$.  The correlation length decreases rapidly ($\rho_l=0.28$) and a passage is noted from normal diffusion to sub-diffusion ($\rho_c=0.22$). From the network analysis this result is understood as a signature of the small-world regime:  strongly connected local clusters are created, which are weakly connected among them, and therefore channels of preferential transport arise, leading to sub-diffusion. 

Numerical results indicate that the combination of structural disorder with nonlinearity leads to almost complete localization of the wavepacket, for values bellow the self-trapping threshold ($\chi_c= 7.5$), which is in agreement with previous investigations~\cite{perakis_discrete_2011,tsironis_exact_2011}. It would be very illuminating to perform the experiments discussed here and actually see whether one can control diffusion through the lattice geometry (disorder) and laser intensity (nonlinearity), and probe experimentally the small-world regime. Additionally, it would be interesting to extend the same line of investigation to different complex network topologies, such as the apollonian networks, in which is known that one can observe quantum phase transitions of light~\cite{halu_2013,halu_2012, bianconi}.

\section{Acknowledgements}
The authors would like to thank Vassilios Kovanis and George Neofotistos for the useful discussions and comments. We  acknowledge partial support through the European Union program FP7-REGPOT-2012-2013-1 under grant agreement 316165.



\begin{thebibliography}{2}
\bibitem{strogatz_exploring_2001}
Steven H. Strogatz Exploring complex networks \textit{Nature} \textbf{410} 268-276 (2001)

\bibitem{vendruscolo_small-world_2002}
M. Vendruscolo, N. V. Dokholyan, E. Paci, M. Karplus Small-world view of the amino acids that play a key role in protein folding, \textit{Phys. Rev. E} \textbf{65} 061910 (2002)

\bibitem{newman_structure_2003}
M. E. J Newman, The Structure and Function of Complex Networks, \textit{{SIAM} Review} \textbf{45} (2003)

\bibitem{rao_protein_2004}
Rao, Francesco and Caflisch, Amedeo The Protein Folding Network, \textit{Journal of Molecular Biology} \textbf{342} 299-306 (2004)
\bibitem{pomerance_effect_2009}
A. Pomerance, E, Ott, M. Girvan,  W. Losert The effect of network topology on the stability of discrete state models of genetic control \textit{W. PNAS} \textbf{106} 8209-8214 (2009)

\bibitem{kempes_predicting_2011}
C. P. Kempes, G. B. West, K. Crowell, M. Girvan Predicting Maximum Tree Heights and Other Traits from Allometric Scaling and Resource Limitations, \textit{PLOS ONE} \textbf{6} e20551 (2011)


\bibitem{watts_collective_1998}
Duncan J. Watts and Steven H. Strogatz  Collective dynamics of `small-world' networks, \textit{Nature} \textbf{393} 440-442 (1998)

\bibitem{liu_controllability_2011}
Y.Y Liu, J.J. Slotine, A.L Barabasi Controllability of complex networks, \textit{Nature} \textbf{473} 167-173 (2011)

\bibitem{halu_2013}
A. Halu, S. Garnerone, A. Vezzani and G. Bianconi,
Phase transition of light in complex networks,
\textit{Phys. Rev. E}  {\bf 87}, 022104 (2013)

\bibitem{halu_2012}
A. Halu, L. Ferretti, A. Vezzani, G. Bianconi,
Phase diagram of the Bose­Hubbard model on complex networks,
\textit{Europhysics Letters} {\bf 99}, 18001 (2012)

\bibitem{bianconi}
G. Bianconi,
Superconductor­insulator transition in annealed complex networks,
\textit{Phys. Rev. E} {\bf 85}, 061113 (2012)

\bibitem{schwartz_transport_2007}
T. Schwartz, G. Bartal S. Fishman, M.Segev, Transport and Anderson localization in disordered two-dimensional photonic lattices \textit{Nature} \textbf{446} 52-55 (2007)

\bibitem{lahini_observation_2009}
Y. Lahini, R. Pugatch, F. Pozzi, M. Sorel, R. Morandotti, N. Davidson, and Y. Silberberg, Observation of a Localization Transition in Quasiperiodic Photonic Lattices, \textit{Phys. Rev. Lett.} \textbf{103} 013901 (2009)

\bibitem{segev_anderson_2013}
M.Segev, Y.Silberberg D.N. Christodoulides, Anderson localization of light \textit{Nat. Photon.} \textbf{7} 197-204 (2013)

\bibitem{naether_enhanced_2013}
U. Naether, S. Rojas-Rojas, A. J. Martínez, S. Stützer, A. Tünnermann, S. Nolte, M. I. Molina, R. A. Vicencio, A. Szameit  Enhanced distribution of a wave-packet in lattices with disorder and nonlinearity, \textit{Opt. Express} \textbf{21} 927-934 (2013)

\bibitem{vardeny_optics_2013}
Z. V. Vardeny,	 A. Nahata	, A. Agrawal, Optics of photonic quasicrystals, \textit{Nat Photon} \textbf{7} 177-187 (2013)

\bibitem{hennig_wave_1999}
D. Hennig G. P. Tsironins, Wave transmission in nonlinear lattices, \textit{Physics Reports} \textbf{307} 333-432 (1999)


\bibitem{li_self-trapping_2013}
S. Li, S. R. Manmana, A. M. Rey, R. Hipolito, A. Reinhard, J.-F. Riou, L. A. Zundel, D. S. Weiss, Self-trapping dynamics in a two-dimensional optical lattice, \textit{Phys. Rev. A} \textbf{88} 023419 (2013)

\bibitem{picozzi_solitons:_2008}
A. Picozzi Solitons: Self-trapping of speckled light beams, \textit{Nat. Phot.} \textbf{2} 334-335 (2008)
\bibitem{gao_iterative_2010}
H. Gao, L. Tian, B. Zhang, G. Barbastathis Iterative nonlinear beam propagation using Hamiltonian ray tracing and Wigner distribution function, \textit{Opt. Lett.} \textbf{35} 4148-4150 (2010)
\bibitem{mattheakis_luneburg_2012}
M. Mattheakis, G. P. Tsironis, V. I. Kovanis, Luneburg lens waveguide networks, \textit{J. Opt.} \textbf{14} 114006 (2012)

\bibitem{molina_dynamics_1993}
M. Molina G. P. Tsironis, Dynamics of self-trapping in the discrete nonlinear Schrödinger equation, \textit{Physica D: Nonlinear Phenomena} \textbf{65} 267-273 (1993)

\bibitem{tsironis_applications_1993}
M. Molina G. P. Tsironis, Applications of self-trapping in optically coupled devices \textit{Physica D: Nonlinear Phenomena} \textbf{68} 135-137 (1993)

\bibitem{molina_disorder_1995}
M. I. Molina, G. P. Tsironis, Disorder in the discrete nonlinear Schrodinger equation, \textit{International Journal of Modern Physics B} \textbf{9} 1899-1932 (1995)

\bibitem{perakis_discrete_2011}
F. Perakis, G. P. Tsironis, Discrete nonlinear Schrödinger equation dynamics in complex networks, \textit{Phys. Lett. A} \textbf{375} 676-679 (2011)

\bibitem{tsironis_exact_2011}
G. P. Tsironis Exact dynamics for fully connected nonlinear networks, \textit{Phys. Lett. A} \textbf{375} 1304-1308 (2011)

\bibitem{ni_origin_2011}
X. Ni, W. Wang, Y. Lai, \textit{Europhys. Lett.} \textbf{96} 44002 (2011)

\bibitem{marte_paraxial_1997}
M. A. M. Marte, S. Stenholm, Paraxial light and atom optics: The optical Schrödinger equation and beyond, \textit{Phys. Rev. A} \textbf{56} 2940-2953 (1997)
\bibitem{barrat_architecture_2004}
A. Barrat, M. Barthelemy, R. Pastor-Satorras, A. Vespignani, The Architecture of Complex Weighted Networks, \textit{PNAS} \textbf{101} 3747-3752 (2004)


\bibitem{callaway_network_2000}
Duncan S. Callaway, M. E. J. Newman, Steven H. Strogatz, and Duncan J. Watts Network Robustness and Fragility: Percolation on Random Graphs, \textit{Phys. Rev. Lett.} \textbf{85} 5468 (2000)

\bibitem{zhao_inducing_2013} 
J.-H. Zhao, H.-J. Zhou, Y.-Y. Liu  Inducing effect on the percolation transition in complex networks \textit{Nat. Commun.}  \textbf{4} 2412 (2013)

\bibitem{eilbeck_discrete_1985}
J. C. Eilbeck, P. S. Lomdahl, A. C. Scott, The discrete self-trapping equation, \textit{Physica D: Nonlinear Phenomena} \textbf{16} 318-338 (1985)



\end{thebibliography}

\end{document}